\documentclass[runningheads]{llncs}

\usepackage{eccv}

\usepackage{eccvabbrv}

\usepackage{graphicx}
\usepackage{booktabs}

\usepackage[accsupp]{axessibility}  
\usepackage{multirow}
\usepackage{multicol}

\usepackage[pagebackref,breaklinks,colorlinks,citecolor=eccvblue]{hyperref}
\usepackage{hyperref}

\usepackage{orcidlink}
\usepackage{float}

\begin{document}

\title{Video-Robin: Autoregressive Diffusion Planning for Intent-Grounded Video-to-Music Generation} 

\titlerunning{Video-Robin}

\author{Vaibhavi Lokegaonkar*\inst{1} \and
Aryan Vijay Bhosale*\inst{1} \and
Vishnu Raj\inst{2} \and
Gouthaman KV\inst{2} \and
Ramani Duraiswami\inst{1} \and
Lie Lu\inst{2} \and
Sreyan Ghosh\inst{1,3} \and
Dinesh Manocha\inst{1}}

\authorrunning{V. Lokegaonkar, A.V. Bhosale et al.}

\institute{University of Maryland College Park, USA \and
Dolby Laboratories, USA \and 
NVIDIA, USA\\
Corresponding authors: \email{\{vlokegao,aryanvib\}@umd.edu}}

\maketitle
\vspace{-1.5em}
\begingroup
\renewcommand\thefootnote{}
\footnotetext{* Equal contribution}
\endgroup

\begin{abstract}
    Video-to-music (V2M) is the fundamental task of creating background music for an input video. Recent V2M models achieve audiovisual alignment by typically relying on visual conditioning alone and provide limited semantic and stylistic controllability to the end user. In this paper, we present \textbf{Video-Robin}, a novel text-conditioned video-to-music generation model that enables fast, high-quality, semantically aligned music generation for video content. To balance musical fidelity and semantic understanding, Video-Robin integrates autoregressive planning with diffusion-based synthesis. Specifically, an autoregressive module models global structure by semantically aligning visual and textual inputs to produce high-level music latents. These latents are subsequently refined into coherent, high-fidelity music using local Diffusion Transformers. By factoring semantically driven planning into diffusion-based synthesis, Video-Robin enables fine-grained creator control without sacrificing audio realism. Our proposed model outperforms baselines that solely accept video input and additional feature conditioned baselines on both in-distribution and out-of-distribution benchmarks with a \textbf{2.21x} speed in inference compared to SOTA. We will open-source everything upon paper acceptance.
\end{abstract}

\section{Introduction}
\label{sec:intro}

\begin{figure}[t]
    \centering
    \includegraphics[width=0.85\linewidth]{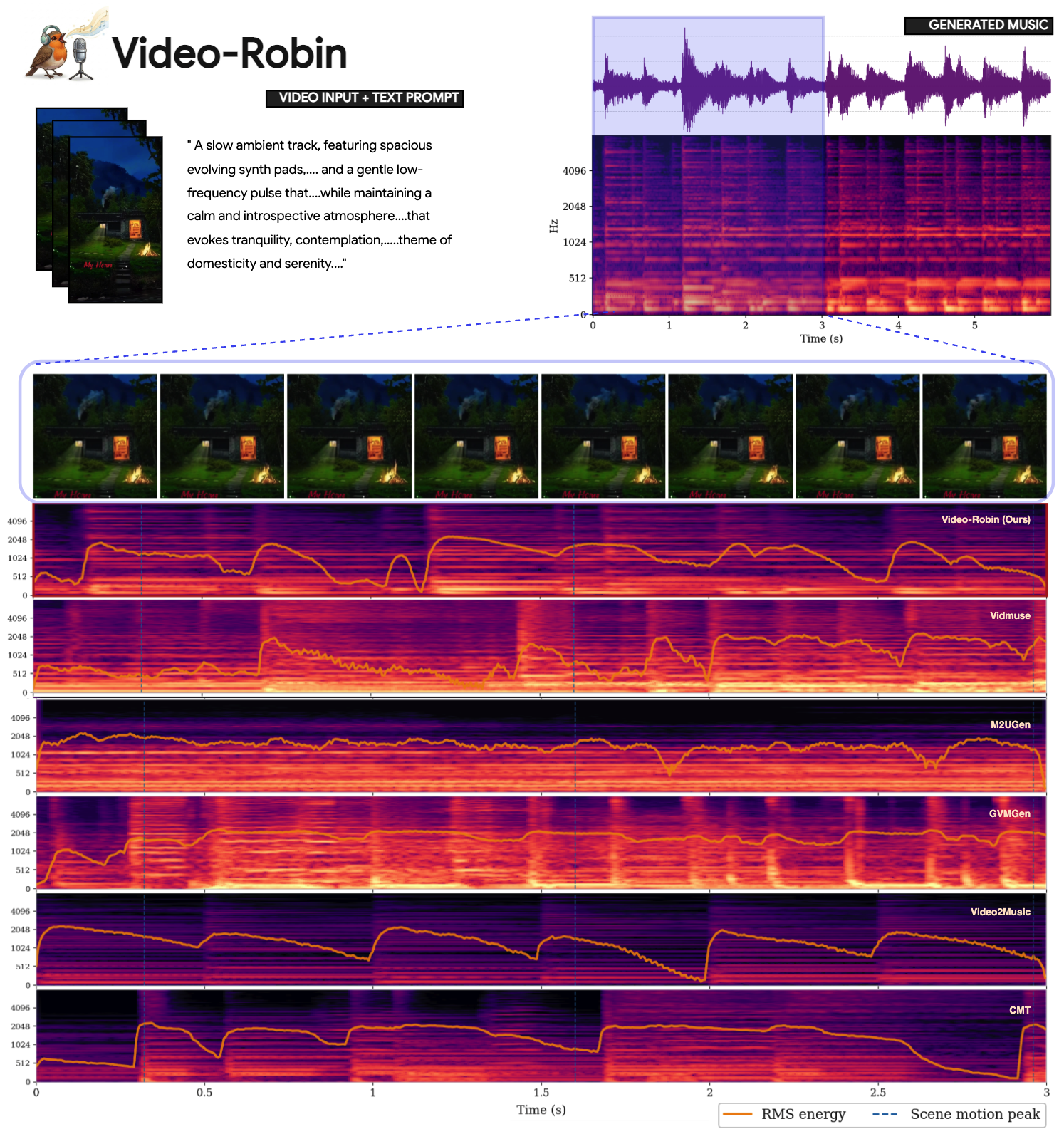}
    \caption{We present \textbf{Video-Robin}, a video$+$text-conditioned music generation model. Above is an example of how Video-Robin takes video frames \& a text prompt as input to generate semantically-aligned music. On zooming in, we see how the generated music adheres faithfully to the nuances of the scenery producing music that aligns to the frame as the fire crackles, smoke rises through the chimney and tress sway in the wind.}
    \label{fig:placeholder}
\end{figure}

\begin{figure}
    \centering
    \begin{subfigure}[b]{0.48\linewidth}
        \centering
        \includegraphics[width=\linewidth]{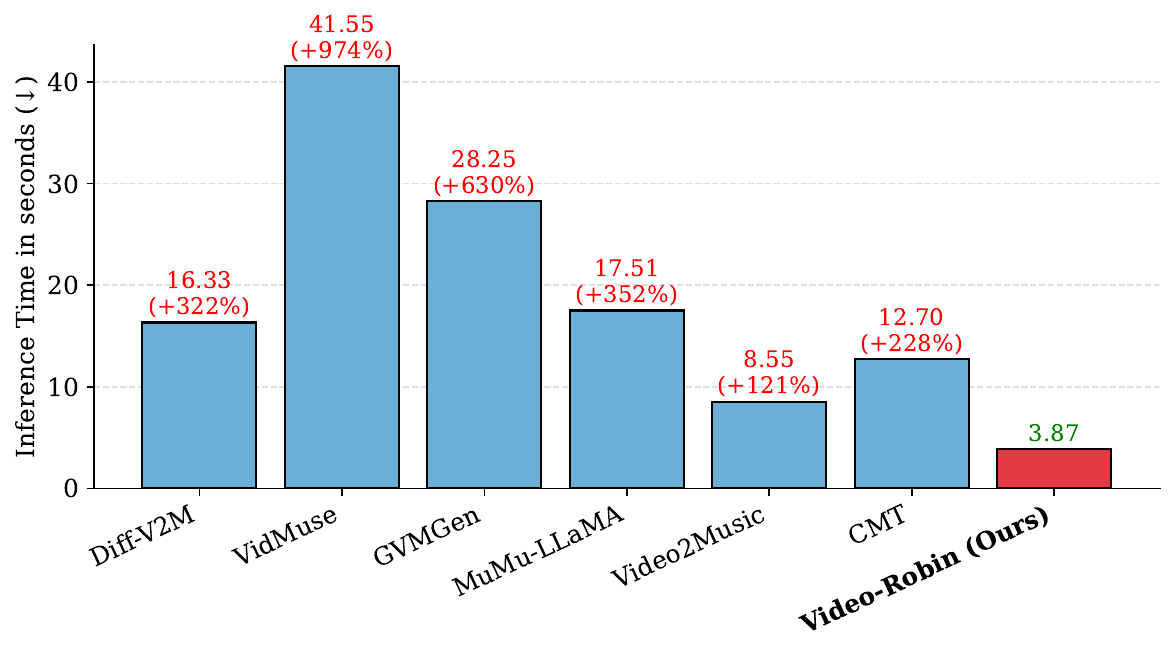}
        \caption{Inference time comparison. Video-Robin is $2.21\times$ faster than the fastest baseline.}
        \label{fig:infer_time}
    \end{subfigure}
    \hfill
    \begin{subfigure}[b]{0.48\linewidth}
        \centering
        \includegraphics[width=\linewidth]{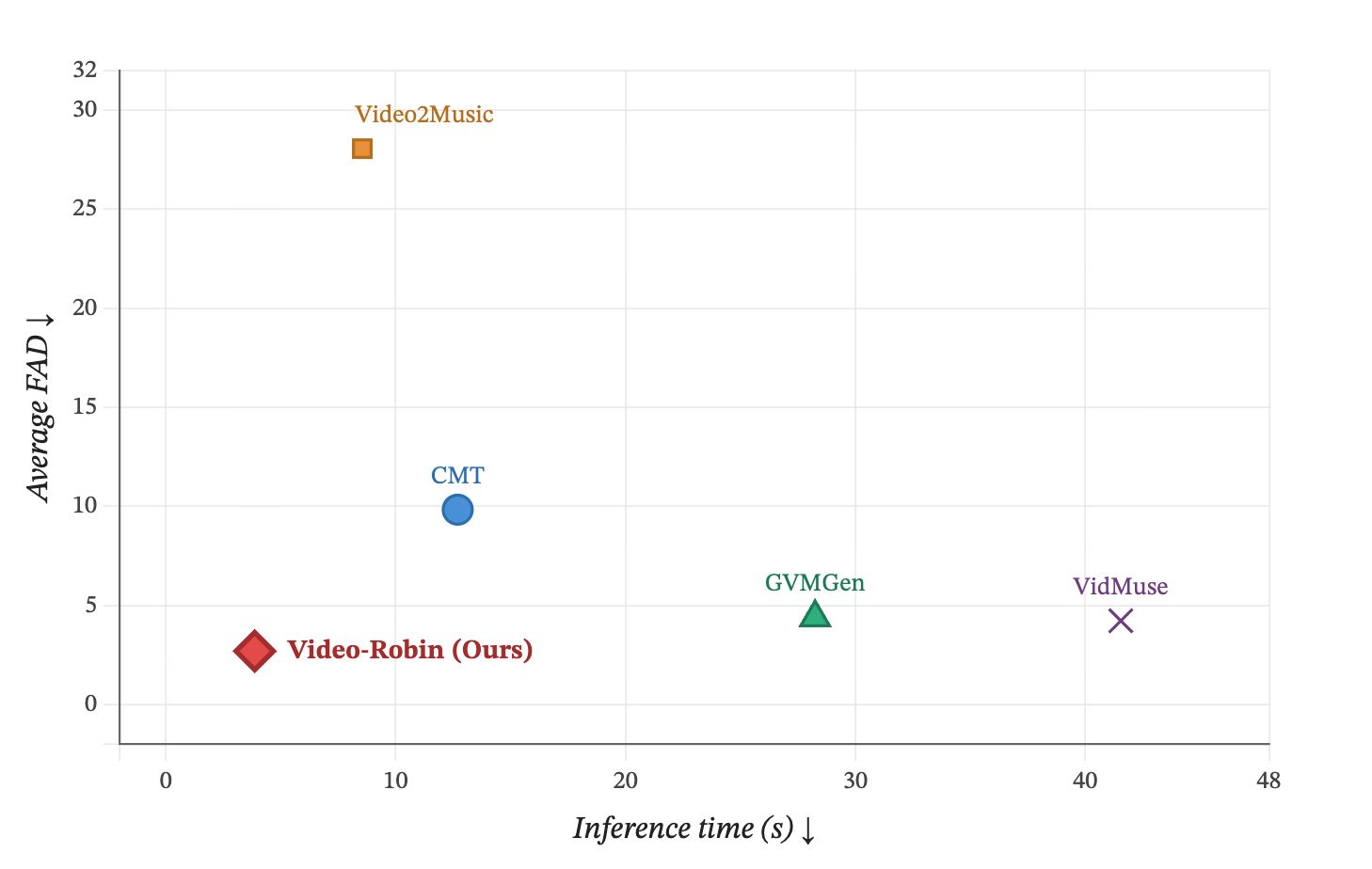}
        \caption{Inference time vs.\ average FAD across ReelBench, LORIS, and V2MBench. Lower is better on both axes.}
        \label{fig:comp}
    \end{subfigure}
    \caption{Efficiency and quality analysis. (a) Video-Robin achieves $2.21\times$ faster inference than the fastest existing baseline. (b) Video-Robin attains the lowest average FAD (2.69) at 3.87s, over $2\times$ faster than Video2Music (8.55s) and $10\times$ faster than VidMuse (41.55s), the only model with comparable audio quality.}
    \label{fig:efficiency}
\end{figure}



Human perception is inherently multimodal, 
enabling music and imagery to jointly shape the narrative, emotion, and memorability of a video. With the ubiquitous creation of short-form videos on social platforms, creators increasingly rely on background scores to boost impact, retention, and shareability, while turning to generative music tools to avoid manual composition and licensing overhead. Yet most current tools prioritise convenience over expressive control, producing plausible but stylistically narrow tracks with limited structural and stylistic steering. As creator-driven content grows in volume and diversity, there is an urgent need for generative systems 
to be avidly expressive for embedding artistic intent into music.\\
Conditional music generation is a well-studied topic of research spanning autoregressive \cite{dhariwal2020jukeboxgenerativemodelmusic, copet2024simplecontrollablemusicgeneration} and diffusion-based \cite{liu2023audioldmtexttoaudiogenerationlatent, liu2024audioldm2learningholistic, evans2024stableaudioopen} paradigms covering various types of conditioning. Earlier models focused on text-to-music generation to produce audio with recognizable musical structure and coherence. As the field matured, the scope of generation expanded beyond instrumental backgrounds to lyrics-to-song synthesis \cite{liu2025songgensinglestageautoregressive, ning2025diffrhythmblazinglyfastembarrassingly} with coordinated vocals and instrumentation. However, autoregressive models capture long-range musical structure effectively but suffer from slow inference and occasional artefacts. Diffusion-based models, on the other hand, offer faster, higher-fidelity synthesis while struggling with global coherence. This has motivated a growing trend toward hybrid, hierarchical architectures \cite{yang2026heartmulafamilyopensourced, gong2025acestepstepmusicgeneration, yang2025songbloomcoherentsonggeneration} that combine the strengths of both. Collectively, these advances demonstrate that text and reference audio-based generation models can produce controllable, temporally coherent music. Despite this progress, the short-form video creator remains underserved. These models produce outputs that are stylistically appropriate but lack the dynamic responsiveness to scene changes, cuts, and pacing shifts that video-first content demands.

Building on prior work, most video-to-music models are built to answer a single question: \textbf{what music fits this video?} Drawing from text-to-music, they use Transformer- and diffusion-based architectures with additions like affective cues \cite{Kang_2024}, long/short-term modeling \cite{tian2025vidmusesimplevideotomusicgeneration}, LLM guidance \cite{liu2024mumullamamultimodalmusicunderstanding}, hierarchical attention \cite{zuo2025gvmgengeneralvideotomusicgeneration}, and explicit rhythm modeling \cite{ji2025diffv2mhierarchicalconditionaldiffusion}, enabling strong audio–visual alignment. But alignment does not always tell the whole story: creators also care about \textit{what music they want} for a video. The same travel vlog may be intentionally scored with ambient electronics or acoustic folk-both “fit,” yet vision-only conditioning cannot express such stylistic and structural intent. Preferences in style, instrumentation, and emotional arc lie outside the current conditioning interface, making it difficult to generate music that matches both the visual narrative and the creator’s artistic goals.

\vspace{0.5mm}
{\noindent \textbf{Main Contributions.}} In this paper, we introduce \textbf{Video-Robin}, a \\text-conditioned video-to-music generation framework that synergistically combines autoregressive and diffusion-based architectures. Inspired by the recent advent of autoregressive Diffusion Transformers in text-to-speech \cite{jia2025ditardiffusiontransformerautoregressive, zhou2025voxcpmtokenizerfreettscontextaware} and text-to-music generation \cite{yang2025songbloomcoherentsonggeneration}, Video-Robin uses an autoregressive planner (AR-Head) coupled with local Diffusion Transformers (Refinement-Head). The AR-Head, featuring a multimodal semantic LM, finite scalar quantization, and a residual integration transformer encoder (RITE), integrates textual and visual inputs with patch-based musical autoregressive embeddings to provide coarser, global latents. The Refinement-Head conditionally denoises these latent patches, which are decoded into high-fidelity music using a VAE Decoder. To summarize, our main contributions are:
\begin{itemize}
    \item We propose Video-Robin, a text-conditioned video-to-music generation framework that combines intent-grounded autoregressive planning with Diffusion Transformer (DiT) refinement to generate controllable, high-quality, semantically aligned music for videos; extensive experiments show it outperforms prior video-to-music methods on both in-domain and out-of-domain benchmarks and formalises text-prompted, video-conditioned music generation as a new task to better capture creators’ stylistic, emotional, and thematic intent.
    
    \item We curate ReelBench, a novel benchmark for text-conditioned video-to-music generation on short-form content. ReelBench comprises 300 samples, each paired with fine-grained generation prompts that specify musical attributes such as key, tempo, and chord progression, spanning diverse emotional categories and thematic domains.
    \item Through extensive experiments we show how our model outperforms state-of-the-art model on generated audio quality, diversity and audio-visual alignment metrics on in-distribution and out-of-distribution datasets with SOTA inference time.

\end{itemize}

\section{Related Works}
\label{sec:related_works}
\subsection{Generative modeling with Language Models for Audio}
Autoregressive models show exceptional performance on understanding tasks while Diffusion-based models excel at generation tasks. Recent works demonstrate how these architectures can be leveraged concurrently for conditional audio generation. ARDiT \cite{liu2024autoregressivediffusiontransformertexttospeech} uses an autoregressive Diffusion Transformer to generate continuous audio tokens for zero-shot text-to-speech and text-based speech editing, DiTAR \cite{jia2025ditardiffusiontransformerautoregressive} builds on this by combining a language model with a diffusion transformer and proposing a patch-based autoregressive framework for high quality speech synthesis. VoxCPM \cite{zhou2025voxcpmtokenizerfreettscontextaware} explored a tokenizer-free approach to the same with semi-discrete residual representations and differentiable quantization and expanded the same to voice cloning. Music generation is a nuanced task for machine learning models, especially when conditioned on multimodal inputs, and can greatly benefit from such hierarchical planning and generation.

\subsection{Text-to-Music Generation Models}
Over the years, we have seen multiple autoregressive and diffusion-based models for text-conditioned music generation. JukeBox \cite{dhariwal2020jukeboxgenerativemodelmusic}. and MusicGen \cite{copet2024simplecontrollablemusicgeneration} pioneered autoregressive music generation with VQ-VAEs and codec-based tokens respectively while AudioLDM \cite{liu2023audioldmtexttoaudiogenerationlatent} and later, AudioLDM 2 \cite{liu2024audioldm2learningholistic} introduced diffusion for general audio tasks, including music generation with improved musical structure and lyrical coherence. Stable Audio Open \cite{evans2024stableaudioopen} utilizes a Diffusion model in an autoencoded latent space, conditioned on T5, to produce longer, high-quality audio. Yue \cite{yuan2025yuescalingopenfoundation} builds on the LLaMa 2 \cite{touvron2023llama2openfoundation} architecture to tackle lyrics-to-song generation through track-decoupled next token prediction and structural progressive conditioning. SongGen \cite{liu2025songgensinglestageautoregressive} employs a single-stage autoregressive Transformer for text-to-song synthesis, emphasizing harmony between vocals and instruments. DiffRhythm \cite{ning2025diffrhythmblazinglyfastembarrassingly} introduces a rhythm-aware latent diffusion framework for temporally coherent and controllable full-song generation. HeartMuLa \cite{yang2026heartmulafamilyopensourced} leverages a hierarchical music LM that autoregressively predicts codec tokens with a global transformer for long-range structure and a local transformer for fine acoustic detail, conditioned on lyrics, tags, and optional reference audio.

Increasingly, we see models adopting a hierarchical approaches for music synthesis. LLM-based models like Yue \cite{yuan2025yuescalingopenfoundation} and SongGen \cite{liu2025songgensinglestageautoregressive} excel at lyric-to-music generation but with slow inference speeds and musical artifacts whereas Diffusion-based architectures enable faster, high-quality generation but struggle with structural coherence. To tackle this, we see ACE-Step \cite{gong2025acestepstepmusicgeneration}integrates DCAE\cite{chen2025dcae15acceleratingdiffusion}-compressed mel-spectrogram latents with flow matching using a linear transformer backbone. SongBloom \cite{yang2025songbloomcoherentsonggeneration} uses a unified autoregressive diffusion architecture with an autoregressive transformer decoder for sketch tokens, a non-autoregressive diffusion transformer for acoustic refinement, and an acoustic encoder, generating interleaved semantic and acoustic patches from lyrics and a short audio reference. 

Inspired by these advances in conditional music generation, Video-Robin has an autoregressive planner to assimilate vision and text signals to produce structured music latents, followed by local Diffusion-transformers to refine these latents into high-fidelity musical output.

\subsection{Video-to-Music Generation Models}
CMT \cite{Di_2021} is one of the earliest works that explores video-to-music generation using a linear Transformer that autoregressively generates symbolic music using compound word representations, conditioning on video through rhythmic feature replacement at inference rather than learned cross-modal alignment. Video2Music \cite{Kang_2024} trains an Affective Multimodal Transformer that takes concatenated video features (semantic, scene offset, motion, emotion) as input to generate music that with emotional alignment. VidMuse \cite{tian2025vidmusesimplevideotomusicgeneration} fuses global and local visual features using Long-Short-Term modeling and generates discrete audio tokens with a Music Token Decoder that can be decoded by an audio codec. MuMu-LLaMA \cite{liu2024mumullamamultimodalmusicunderstanding} uses multi-modal encoders (MERT \cite{li2024mertacousticmusicunderstanding}, ViT \cite{dosovitskiy2021imageworth16x16words}, ViViT \cite{arnab2021vivitvideovisiontransformer}) coupled via adapters to LLaMA 2 \cite{touvron2023llama2openfoundation} to output conditioning embeddings for AudioLDM 2 \cite{liu2024audioldm2learningholistic} or MusicGen\cite{copet2024simplecontrollablemusicgeneration}-decoded music generation. GVMGen \cite{zuo2025gvmgengeneralvideotomusicgeneration}, on the other hand, explores hierarchical spatial and temporal attentions on hidden features to enforce audio-visual alignment. DiffV2M \cite{ji2025diffv2mhierarchicalconditionaldiffusion} utilizes explicit rhythmic modeling, emotional and semantic feature extraction from videos for conditional music generation built on a DiT-based latent diffusion model. Unlike prior models that depend predominantly on visual signals, Video-Robin explicitly incorporates textual intent through autoregressive planning and diffusion refinement enabling finer creative control.

\section{Dataset}
\label{sec:dataset}

\subsection{ReelBench}
\begin{figure}[t]
    \centering
    \begin{subfigure}[b]{\linewidth}
        \centering
        \includegraphics[width=0.8\linewidth]{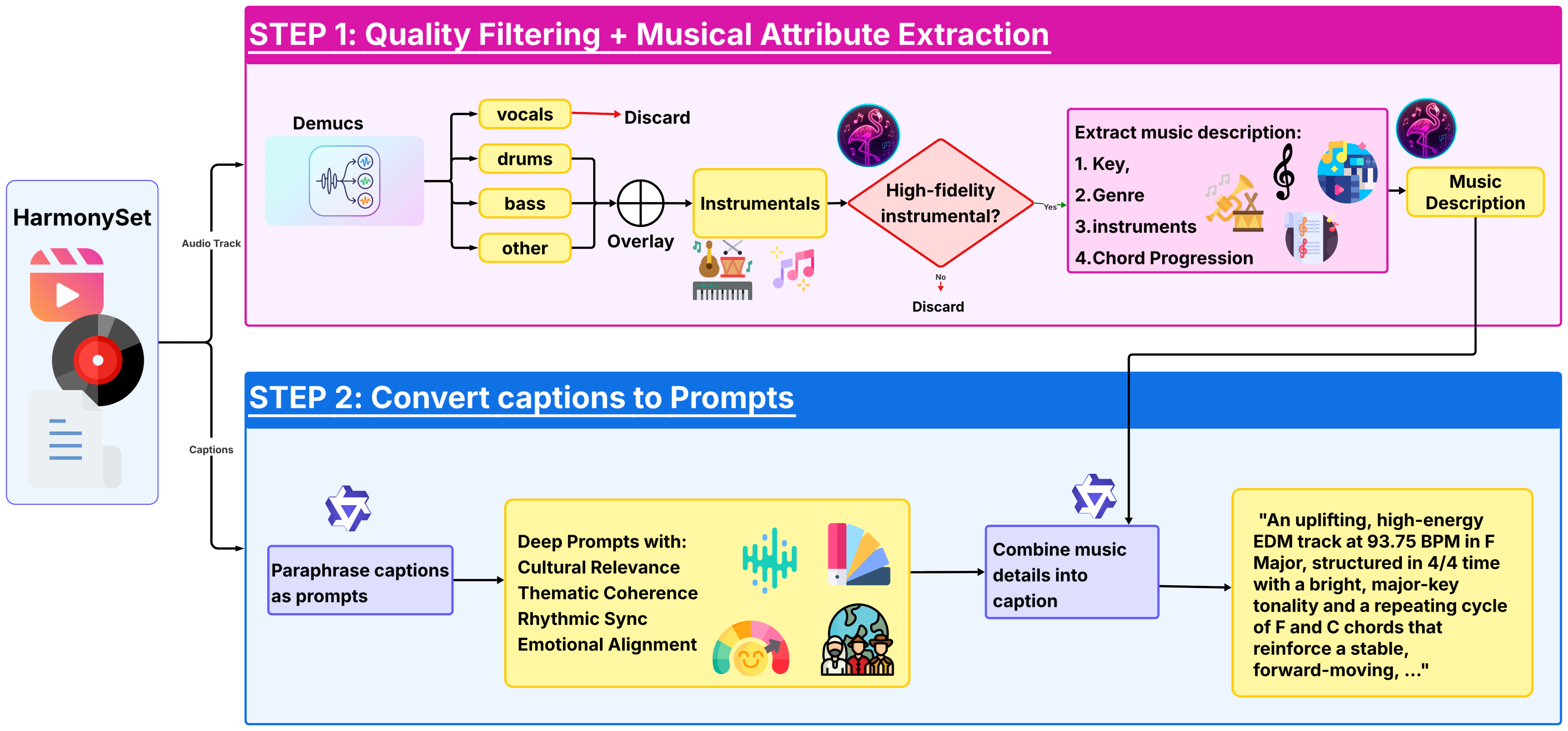}
        \caption{Data preprocessing pipeline.}
        \label{fig:dataprep}
    \end{subfigure}
    
    \vspace{0.5em}
    
    \begin{subfigure}[b]{\linewidth}
        \centering
        \includegraphics[width=0.8\linewidth]{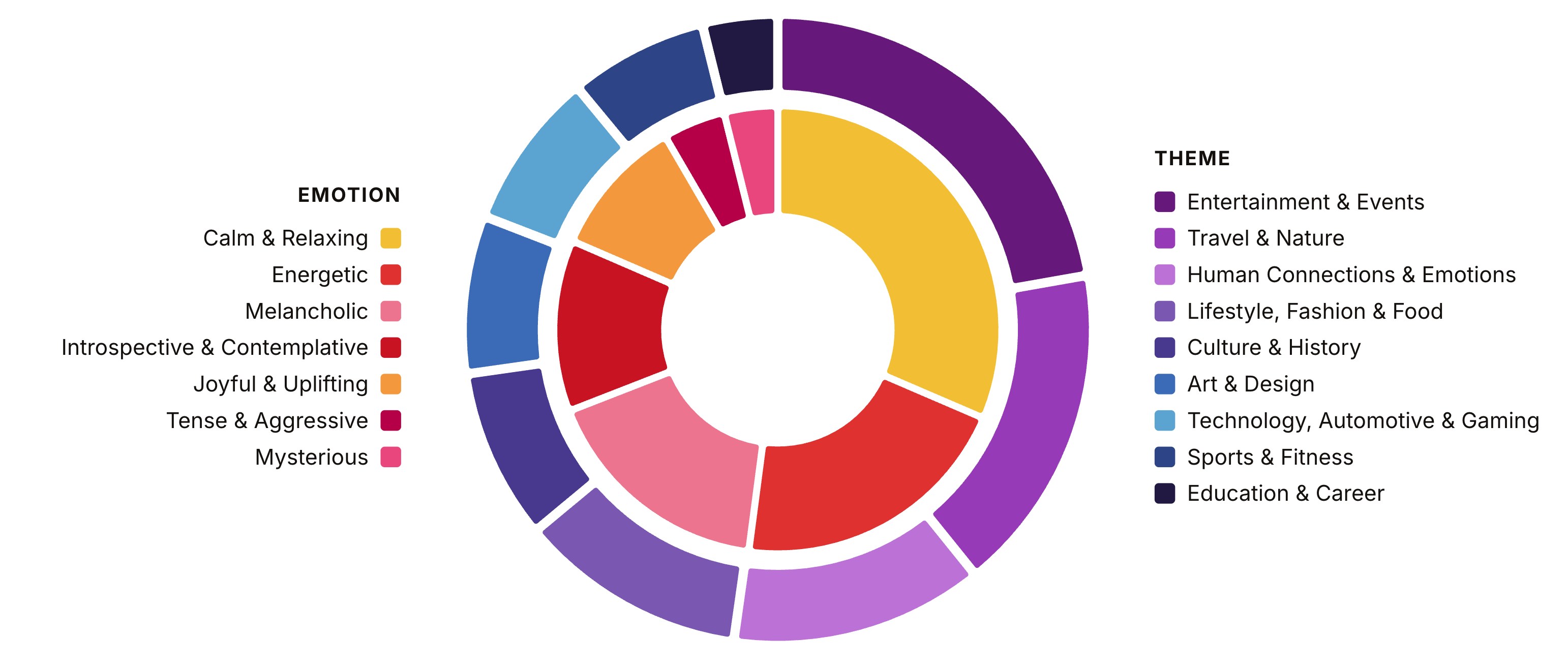}
        \caption{ReelBench data distribution over emotions and themes.}
        \label{fig:reelbench_dist}
    \end{subfigure}
    
    \caption{(a) Overview of the data preprocessing pipeline used to construct fine-grained training prompts from HarmonySet captions and MusicFlamingo-extracted musical attributes. (b) Distribution of ReelBench across emotion and theme categories.}
    \label{fig:dataset_overview}
\end{figure}

In the previous section we establish how the evolving landscape of video-to-music generation and now, text$+$video-to-music generation suffers from lack of a comprehensive dataset that can allow for fine-grained stylistic, thematic and structural control through text prompting. To tackle this, we curate \textbf{ReelBench}, a comprehensive evaluation benchmark for text$+$video-to-music generation. Reelbench builds on the abundancy of paired audio-video data in video-understanding models through datasets like HarmonySet \cite{zhou2025harmonysetcomprehensivedatasetunderstanding}. These datasets contain metadata about audio-visual alignment but lack the structure, instruction and fine-grained detail required for generation tasks. We adopt a comprehensive pipeline with human verification to produce rich prompts for highly controllable music generation using Music Flamingo \cite{ghosh2025musicflamingoscalingmusic}, a music understanding model and Large Language Models (LLMs), leveraging the already available metadata to ground our prompts. Figure \ref{fig:reelbench_dist} shows the distribution of data over various emotion and themes establishing the diversity of the data. We release ReelBench with \textbf{300 data points}, each with high-quality audio, video and generation prompts.

\subsection{Data Preprocessing}
We employ the HarmonySet \cite{zhou2025harmonysetcomprehensivedatasetunderstanding} dataset for both training and evaluation. HarmonySet is a video-to-music alignment dataset consisting of short-form videos with rapid scene transitions and contemporary music tracks, reflecting the dynamic nature of social media content. Our task focuses on generating background music that aligns both temporally and semantically with such fast-changing visuals. Each video in HarmonySet is accompanied by human-reviewed deep captions covering four alignment dimensions: Rhythmic Synchronization, Thematic Coherence, Emotional Alignment, and Cultural Relevance. These captions provide fine-grained markers that guide the model to generate diverse and contextually relevant music.

We first separate audio from the visual content from each video in Harmony set and apply Demucs to separate the instrumentals and vocals. Demucs separates the audio into four stems: \textit{vocals}, \textit{bass}, \textit{drums} and \textit{other}. We obtain the instrumentals by overlaying the \textit{bass}, \textit{drums} and \textit{other} stems. We then utilise MusicFlamingo \cite{ghosh2025musicflamingoscalingmusic} to estimate whether the instrumental only has well-formed, high-fidelity music without any noisy artefacts. In order to obtain textual prompts with fine-grained details, for each data sample with high fidelity instrumentals, we leverage MusicFlamingo to extract detailed musical attributes in the form of a textual description from the instrumentals, including keys, genre, instruments and chord progression. We then use Qwen3-8B \cite{yang2025qwen3technicalreport} to paraphrase HarmonySet captions into prompts, ensuring that the prompt contains cultural, rhythmic, thematic, and emotional information. These multi-faceted prompts are then combined with the extracted music description using Qwen3-8B to produce prompts that contain a holistic description of the instrumental music and how it aligns with the video. This approach enables our model to generate music that is not only rhythmically synchronised but also semantically coherent with complex visual content. We attach all input prompts to MusicFlamingo and Qwen-3B  in Appendix B.

\section{Methodology}
\label{sec:method}
\begin{figure}
    \centering
    \includegraphics[width=0.83\linewidth]{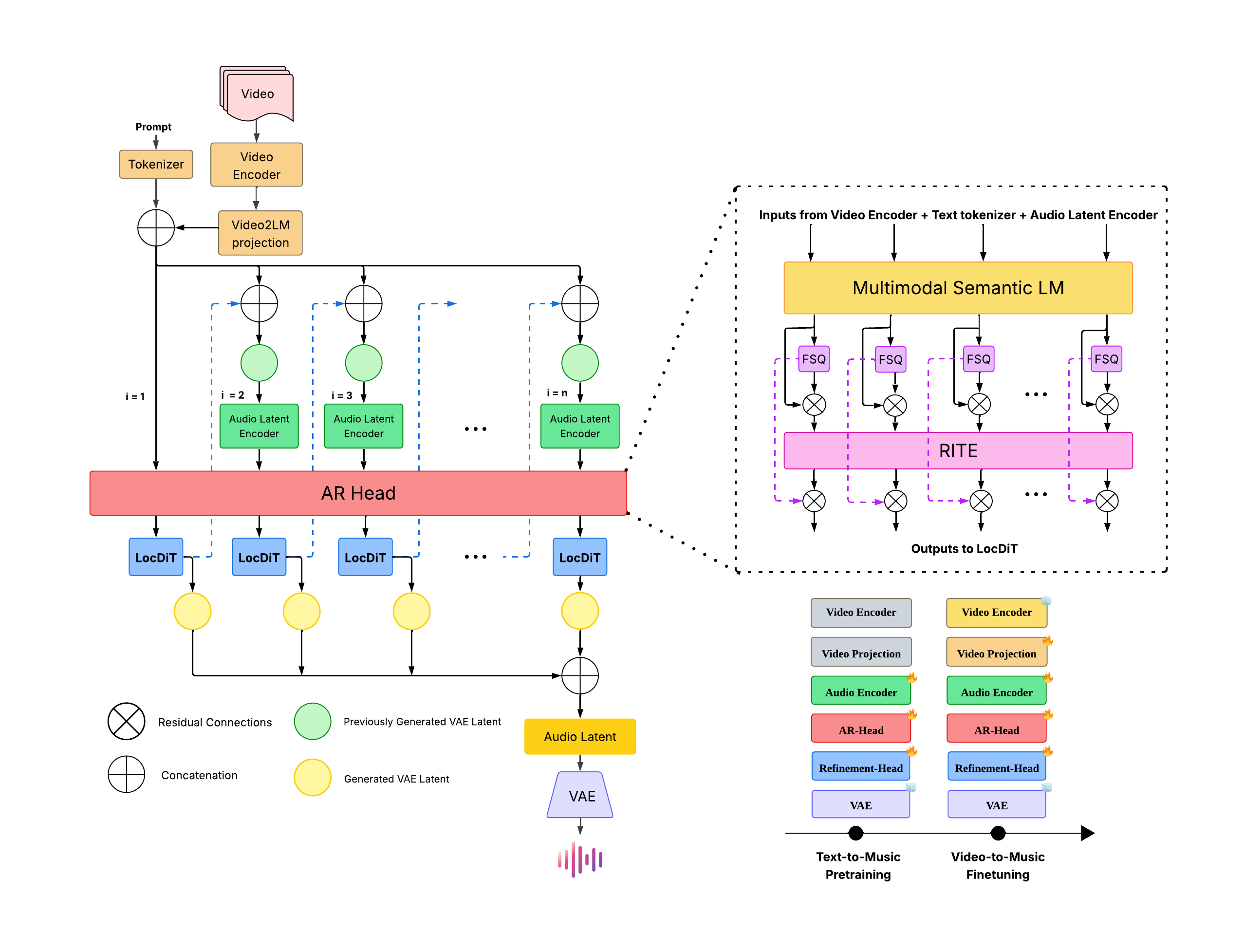}
    \caption{\small \textbf{Overview of the Robin Architecture.} Video and text are autoregressively decoded into VAE latent patches via AR-Head planning and Loc-DiT denoising, then reconstructed into music by the VAE decoder. Training is staged: first aligning text with audio, then learning a projection for video-conditioned music generation.}
    \label{fig:arch}
\end{figure}

Inspired by the success of autoregressive diffusion models in text-to-speech (TTS) \cite{jia2025ditardiffusiontransformerautoregressive}, \cite{zhou2025voxcpmtokenizerfreettscontextaware}, we adapt this paradigm for music generation conditioned on video and fine-grained textual descriptions. The following subsections detail our problem statement and architecture, illustrated in Figure \ref{fig:arch}. 

\subsection{Problem Definition}
Consider an input video $V$ with $t$ frames, $c$ channels, height $h$, and width $w$, and a textual description $T$ with $l$ tokens, such that $V \in \mathbf{R}^{t \times c \times h \times w}$ and $T \in \mathbf{R}^{l}$. We use $V$ and $T$ as conditioning inputs to generate a set of $n$ music latent patches $M = (m_1, \dots, m_n)$ which can be concatenated and decoded to reconstruct the full music waveform. Each patch has length $p$ and $k$ channels, such that $M \in \mathbf{R}^{n \times p \times k}$. Our objective is to learn a model that estimates the probability of the sequence of latent patches $M$ conditional on $V$ and $T$, given by $p_{\theta}(M \mid V, T)$, where $\theta$ are the parameters of the model. We decompose the joint probability of the sequence to be predicted as follows: 

\begin{equation}
    p_{\theta}(M \mid V, T) = \prod_{j= 1}^{n} p_{\theta}(m_j \mid m_1, \dots, m_{j-1}, V, T)
\end{equation}

\subsection{Architecture}

\textbf{Motivation.} The goal of this work is to generate high-fidelity music that is temporally aligned with video and consistent with detailed textual descriptions. Existing approaches rely on discrete codebooks, which limit audio fidelity and face scalability challenges as codebook size or latent dimensionality increases. Directly generating continuous latent sequences avoids these limitations but can overwhelm the model when conditioning on detailed text, risking loss of semantic alignment with video and text. To address this, we adopt an autoregressive framework over continuous latent patches, preserving both acoustic fidelity and alignment with the input modalities.\\

\textbf{Overview.} We propose an autoregressive architecture to model the sequence of music latent patches. Video frames are first encoded using a visual encoder to obtain framewise embeddings. Audio latent generation is decomposed into two components: the \textbf{AR-Head} and \textbf{Refinement-Head}. The AR-Head takes the visual and text embeddings, along with encoded representations of previously generated latent patches from the Refinement-Head, and produces a hidden representation for the current latent patch. The Refinement-Head uses this hidden representation and the embedding of the last generated patch to generate the latent patch in the Variational Autoencoder’s (VAE) latent space. The generated patch is then embedded using an Audio Latent Encoder and then fed back to the AR-Head to condition the generation of the subsequent patch. Once all patches are generated, they are concatenated, and the VAE decoder reconstructs the high-fidelity musical waveform using this concatenated vector.  All components, except the VAE, are trained end-to-end to ensure representation consistency between the autoregressive prior and the refinement module.\\ \\
\textbf{Visual Encoder.} We first embed the video frames $V \in \mathbf{R}^{t \times c \times h \times w}$ using CLIP's~\cite{radford2021learningtransferablevisualmodels} Visual Encoder to obtain framewise visual features $f_{clip} \in \mathbf{R}^{t \times d_v}$, where $d_v$ is the embedding dimension of the visual encoder. These embeddings are then projected into the text embedding space of the AR-Head, $f_v \in \mathbf{R}^{t \times d}$, where $d$ is the embedding dimension of our model, via a trainable linear layer. \\ \\
\textbf{Audio Latent Encoder.} It is a transformer encoder that extracts audio features $f_a \in \mathbf{R}^{(i-1) \times d}$ that captures embedding  from the input VAE latent patches, where $d$ is the dimension of our model. We pass the previously generated latent patches in $(m_1 \dots m_{i-1})$ to an Audio Latent Encoder.  The audio features represent historical context in the form of concise audio embeddings. \\ \\
\textbf{AR-Head.} The AR-Head is responsible for providing a coarse latent description of the audio being generated through the current patch. To circumvent the challenge of tasking one transformer module with all such nuances, we decompose this module further into three components: the Multimodal Semantic LM, the Finite Scalar Quantization layer, and the Residual Integration Transformer. We discuss each of these modules in detail below: \\ 
\noindent \textbf{Multimodal Semantic LM (SemanticLM):} 
    The SemanticLM layer is a transformer encoder responsible for integrating textual, visual, and autoregressive patch embeddings to produce a coarse semantic plan for the current latent patch. Specifically, the framewise projected visual embeddings $f_v$, text tokens $T$, and embeddings of previously generated patches $f_a$ are provided as inputs to this layer for generating an integrated multimodal embedding. The SemanticLM first embeds the text tokens into vector representations, $f_t \in \mathbf{R}^{l \times d}$, where $d$ is the embedding dimension of our model and $l$ is the number of text tokens. The visual, textual and historical patch embeddings are then concatenated and processed to obtain a semantic embedding $E_s \in \mathbf{R}^{(l + t + i - 1) \times d}$.  While doing so, it explicitly captures the relationships between neighbouring patches and the conditioning inputs, as shown below. \\
\noindent \textbf{Finite Scalar Quantization Layer (FSQ):} The FSQ layer acts as a structured bottleneck that promotes stable, high-level semantic representations, improving autoregressive consistency and enabling effective separation between semantic planning and continuous latent refinement. We pass the obtained semantic embedding $E_s$ from the previous component to this differentiable Finite Scalar Quantization (FSQ) layer to obtain semi-discrete embeddings $E_d \in \mathbf{R}^{(l + t + i - 1) \times d}$. 

    \begin{equation}
        E_d = \text{FSQ(\text{SemanticLM(Concat($f_v$, $f_t$, $f_a$))})} = \Delta \cdot \operatorname{clip}\!\left(
    \operatorname{round}\!\left(
        \frac{E_s}{\Delta}
    \right),
    -L, L
\right)
\end{equation}
\\
\noindent \textbf{Residual Integration Transformer Encoder (RITE):} RITE complements the FSQ bottleneck by modeling residual information necessary for acoustic fidelity. While FSQ enforces stable semantic structure, this transformer refines these representations to recover fine-grained details, producing a comprehensive planning embedding for accurate latent generation. The semi-discrete embeddings $E_d$ are further processed by this transformer layer to integrate the residual information not captured by the FSQ bottleneck into the coarse latent description that is sent to the Refinement-Head.
  We finally add the semi-discrete embeddings obtained from the FSQ layer to the embeddings obtained from RITE. This combined representation, referred to as the planning embedding, $E_p \in \mathbf{R}^{(l + t + i - 1) \times d}$ encodes the semantic plan for generating the $i^{th}$ latent patch conditioned on multimodal context and autoregressive history. Effectively, the following computations are done to obtain the planning embedding from the AR-Head:
\begin{equation}
    E_p = E_d + \text{RITE($E_d$)}
\end{equation}
\newline
\textbf{Refinement-Head (LocDiT).} The Refinement-Head is analogous to the LocDiT module in the DiTAR architecture~\cite{jia2025ditardiffusiontransformerautoregressive}. It is implemented as a Diffusion Transformer (DiT) that generates the $i^{th}$ latent patch conditioned on the planning embedding $E_p$, and the previously generated latent patch $m_{i - 1}$. At each generation step of our autoregressive model, the Refinement-Head applies bi-directional attention to denoise the latent patch, refining fine-grained acoustic details guided by the coarse semantic plan provided by the AR-Head. Specifically the following equation describes the production of the $i^{th}$ latent patch:
\begin{equation}
\label{eq:refinement}
m_i = \text{LocDiT}\!\left(
x \sim \mathcal{N}(0, I),\;
\text{Concat}\!\left(E_p,\,\; m_{i-1}\right)
\right)
\end{equation}
This connects the autoregressive generation loop back to Equation~\ref{eq:refinement}: each call to the Refinement-Head runs a full diffusion trajectory, producing a clean latent patch $m_i$ that is then encoded by the Audio Latent Encoder and fed back into the AR-Head for the next patch. \\ \\
\textbf{Causal Variational Autoencoder (VAE).} Our model operates in the latent space of a pretrained VAE from SongBloom, which remains frozen throughout training. Ground-truth waveforms are encoded into latent patches, and during inference, patches generated by the Refinement-Head are decoded and concatenated to produce a high-fidelity waveform temporally aligned with the video and consistent with the text prompt.

\subsection{Training Loss} 
We use a flow-matching diffusion loss to optimise the ODE solver in the LocDiT in the Refinement-Head. The following equation describes this flow-matching loss, where $x_t = \alpha t x_0 + \sigma_t \epsilon$ is the noisy latent at the step $t$ of the denoising process with $\epsilon \sim \mathcal{N}(0, I)$ and $v_{\theta}$ is the velocity field predicted by the LocDiT: 

\begin{equation}
    \mathcal{L}_{\text{diff}} = \mathbb{E}_{t, \mathbf{x}^0, \boldsymbol{\epsilon}}
    \Bigg[
    \Big\|
    \mathbf{v}_\theta(\mathbf{x}^t, \mathbf{E_p}, \mathbf{m}_{i-1}) 
    - \frac{d}{dt} \big( \alpha_t \mathbf{x}^0 + \sigma_t \boldsymbol{\epsilon} \big)
    \Big\|_2^2
    \Bigg]
\end{equation}

\subsection{Multi-Stage Training}
To ensure that the model can follow fine-grained textual prompts, it must learn strong correspondences between musical concepts (e.g., key, tempo, and chord progression) and their realizations in the latent audio space. We therefore adopt a two-stage training strategy. \\
\textbf{Stage 1: Text-to-Music Pretraining.} 
We first train the architecture for text-conditioned music generation by removing the video encoder and projection layer. This stage learns to map structured musical descriptions into latent representations that decode into high-fidelity audio, establishing alignment between textual musical attributes and their acoustic realizations. \\
\textbf{Stage 2: Video-to-Music Finetuning.} 
We then introduce the video encoder and a trainable projection layer that maps visual features into the multimodal conditioning space. The video encoder is kept frozen, while the projection layer is trained from scratch and the remaining model parameters are fine-tuned from the pretrained checkpoint. This initialization stabilizes optimization and enables faster convergence for video-aligned music generation. \\

\section{Experiments}
\label{sec:expt}
\subsection{Datasets}
\textbf{Training Data.} In the pre-training stage, we use the JamendoMaxCaps dataset~\cite{roy2025jamendomaxcaps} to train the model to generate music conditioned on fine-grained text prompts. Our dataset comprises $\approx$1.6M music samples (all instrumental; no vocals), each paired with captions and with an average duration of 30 seconds each. Next, we fine-tune pretrained checkpoints along with the video encoder and projection layers on the training split of the Harmony Set~\cite{zhou2025harmonysetcomprehensivedatasetunderstanding}. After applying the preprocessing pipeline described in Section~\ref{sec:dataset}, the resulting training set consists of 112k video–background music pairs accompanied by fine-grained textual prompts. Each video is 10 seconds long, and all audio is standardized to 48 kHz stereo. The prompts provide detailed musical attributes, including tempo, key, and chord progression, allowing the model to learn structured text-to-music alignment in addition to audiovisual correspondence. \\
\textbf{Evaluation Benchmarks.} To assess both in-distribution performance and generalisation, we evaluate on multiple benchmarks. ReelBench (Section~\ref{sec:dataset}) serves as the in-distribution benchmark, as it is derived from the test split of Harmony Set and shares similar visual dynamics and prompt structure while remaining strictly disjoint from the training set. In contrast, V2MBench~\cite{tian2025vidmusesimplevideotomusicgeneration} and the test split of LORIS~\cite{Yu2023Long} are treated as out-of-distribution benchmarks. These datasets differ from the Harmony Set in terms of visual genre, recording conditions, and the absence of detailed textual prompts. Moreover, since V2MBench and LORIS do not provide prompt annotations, they allow direct comparison with video-only music generation baselines.\\
\textbf{Preprocessing \& Standardization.} For all evaluation datasets, we remove vocal components from the ground-truth audio and verify the quality of the separated instrumental tracks using MusicFlamingo~\cite{ghosh2025musicflamingoscalingmusic}. We discard samples that contain artefacts or non-musical content. To ensure consistency across benchmarks, all videos are truncated to 10 seconds to match our fixed-length generation setting. For datasets without textual prompts, we provide a static instruction prompt: \textit{Generate aligned music for the video}.


\subsection{Implementation Details} 
Our model is built upon the VoxCPM framework and operates in the latent space of a pretrained SongBloom audio VAE at 48\,kHz, which remains frozen during all stages of the training process. The SemanticLM backbone is initialised from MiniCPM (0.5B) and comprises 24 transformer layers with a hidden dimension of 1024 and 16 attention heads. The autoregressive planning head consists of a finite scalar quantization (FSQ) bottleneck with latent dimension 256, followed by RITE, an 8-layer transformer for discrete semantic token modeling. The refinement head, architecturally analogous to LocDiT~\cite{jia2025ditardiffusiontransformerautoregressive}, is implemented as an 8-layer Diffusion Transformer trained with conditional flow matching using log-normal timestep sampling. We use the CLIP-ViT-Base model with a patch size of 32 as the vision encoder. 

We pre-train the model without the video encoder and projection layers on the text-to-music task for 120K steps. The batch size for training is limited to 8, and we employ a learning rate of $10^{-3}$. Our model is trained on 64 H100s. We then use this pretrained checkpoint along with the frozen video encoder and learnable linear projection layer to train all our ablations for the video-to-music task. We train each video-to-music model for 4 epochs using AdamW with weight decay 0.01 and a cosine learning rate schedule with 10\% warmup and peak learning rate $1\times10^{-4}$. Training is conducted on 8 NVIDIA RTX A6000 GPUs for approximately two days. During inference, we integrate the learned velocity field using an Euler solver with 20 diffusion steps and apply classifier-free guidance with a scale of 2.0.

\subsection{Evaluation Metrics}
We quantitatively evaluate the effectiveness of our model using previously adopted metrics for music generation. These metrics evaluate the quality, fidelity and diversity of the generated music. We use \textbf{Frechet Audio Distance (FAD)}, \textbf{Frechet Distance (FD)}, \textbf{KL Divergence (KL)} \cite{zhang2023propertieskullbackleiblerdivergencemultivariate}, \textbf{Inception Score (IS)} \cite{salimans2016improvedtechniquestraininggans}, \textbf{ImageBind Score (IB)} \cite{girdhar2023imagebindembeddingspacebind}, \textbf{Density}, and \textbf{Coverage} \cite{ferjad2020icml}. The quality of the generated music is measured by FAD, FD, and KL, while its alignment with the video is measured by the ImageBind Score. Since ImageBind is not specifically trained on music data, we also employ Gemini as an Omni-Judge to evaluate the video-to-music alignment based on 7 broad axes: rhythmic sync, theme coherence, emotion alignment, cultural relevance, temporal dynamics,  instrumention fit and overall alignment. We attach the prompt administered to Gemini for the evaluation in Appendix A.

\subsection{Comparison Models}
Prior approaches to video-conditioned music generation rarely incorporate fine-grained textual conditioning for controllable synthesis. Since our framework explicitly supports structured text controls, we evaluate performance under two settings: (i) video-conditioned music generation with fine-grained prompts and (ii) video-only conditioning without textual inputs. We compare against \textbf{CMT}~\cite{Di_2021}, \textbf{Video2Music}~\cite{Kang_2024}, and \textbf{M$^2$UGen}~\cite{liu2024mumullamamultimodalmusicunderstanding} as primary baselines in the text-conditioned setting, as these methods incorporate auxiliary textual signals. In the video-only setting, we compare against \textbf{VidMuse}~\cite{tian2025vidmusesimplevideotomusicgeneration} and \textbf{GVMGen}~\cite{zuo2025gvmgengeneralvideotomusicgeneration}, which perform music generation conditioned solely on visual features. For fair comparison, we additionally report our model’s performance without textual inputs to align with these baselines.

\subsection{Quantitative Comparison against baselines}
Across the quality-focused metrics (FAD, FD, and KL), Video-Robin consistently achieves the lowest or near-lowest scores, reflecting superior audio fidelity and distributional alignment with ground-truth music. On ReelBench, Video-Robin attains the best FAD, FD, and Coverage alongside the highest Inception Score, demonstrating that our model generates music of exceptional perceptual quality and diversity. On the LORIS dataset, Video-Robin again leads on FAD, IS, Density, and Coverage, indicating that our generations are not only realistic but also diverse and well-distributed relative to the reference set. On V2MBench, while VidMuse achieves the best scores on IB, Density, and Coverage, Video-Robin remains highly competitive with a strong IS and notably low KL divergence, suggesting that our model generalizes well across diverse video domains. Taken together, these results demonstrate that Video-Robin strikes a favorable balance across audio quality, semantic alignment, and generative diversity, outperforming all baselines on the majority of metrics across benchmarks.  
\subsection{Effect of Removing FSQ and RITE} Table~\ref{tab:ablation-fsq-rite} presents ablation experiments that isolate the contributions of the Finite Scalar Quantization layer (FSQ)
and the Residual Integration Transformer Encoder (RITE) by sequentially removing each component. We evaluate three configurations: the full Video-Robin model, a variant retaining FSQ but removing RITE (\textit{w/o RITE}), and a variant removing both modules (\textit{w/o FSQ + RITE}), which operates entirely on continuous latents. Results are reported across ReelBench, LORIS, and V2MBench to assess generalisability.\\
\textbf{FSQ without RITE degrades performance most severely.}
Retaining the quantization bottleneck while removing the residual
encoder yields the worst results across nearly all metrics and datasets. On ReelBench and LORIS, FAD and FD increase, while the Inception Score collapses across all benchmarks, and diversity metrics drop sharply. This pattern suggests that FSQ, by
design, discards fine-grained information from the semantic embeddings, and without RITE to recover these lost details, the planning embeddings become too coarse for the Refinement-Head to produce high-fidelity or diverse outputs.\\
\textbf{Removing both FSQ and RITE is less harmful than FSQ alone.}
The continuous-latent variant (\textit{w/o FSQ + RITE}) consistently outperforms the FSQ-only variant (\textit{w/o RITE}). For instance, on V2MBench, FAD and Coverage improve when both modules are removed compared to removing only RITE. This confirms that the quantization bottleneck without a residual recovery mechanism actively harms generation quality. Additionally, continuous latents, while lacking the structural benefits of semi-discretisation, at least preserve the information needed for reasonable synthesis.\\
\textbf{The full model demonstrates clear symbiosis between FSQ and
RITE.} When both modules are present, Video-Robin substantially
outperforms both ablated variants on audio fidelity (FAD, FD),
perceptual quality (IS), and generative diversity (Density, Coverage) across all three datasets. The full model achieves the highest Inception Scores by a wide margin and the best Coverage scores. We attribute this to the complementary roles of the two modules: FSQ enforces a structured, stable semantic bottleneck that regularises the planning space, while RITE recovers the residual acoustic detail lost during quantization. Together, they produce planning embeddings that are both semantically coherent and informationally complete, enabling the Refinement-Head to generate diverse, high-quality music. We note that KL divergence is the one metric where ablated variants occasionally achieve lower scores (e.g., \textit{w/o RITE} on LORIS and V2MBench), likely because the degraded outputs collapse toward a narrower distributional mode that happens to align with the reference distribution on this measure; this does not reflect improved generation quality, as all other metrics confirm substantial degradation.
\subsection{Effect of Patch Size on model performance} Additionally, we investigate the effect of spatial patch granularity on generation quality by varying the patch size (4, 8, 16) across different benchmark datasets (Table \ref{tab:ablation-patch}). A recurring trend across metrics is that smaller patch sizes yield superior perceptual fidelity. Audio fidelity metrics (such as FAD \& FD) see monotonic improvement with smaller patches. However, diversity-based metrics (such as KL-Divergence), which depend on the model's semantic understanding, see non-monotonic trends and benefit from larger patches. Additionally, we see how larger patch sizes show similar performance with video-music alignment metrics such as the ImageBind score.  
\subsection{Inference time comparisons} 
We evaluate all baselines on V2MBench using the preprocessing pipeline described above, producing 10-second video inputs for each model. Under this standardized setup, Video-Robin achieves substantially lower inference time than all competing methods. Specifically, it is $2.21$ times faster than the strongest baseline, Video2Music, which is the fastest prior approach. Importantly, this efficiency is achieved without compromising quality: Video-Robin surpasses both text- and video-conditioned baselines in audio fidelity and audio–visual alignment. We attribute the improved inference speed to the reduced number of model parameters and the use of the flow matching objective with a Euler solver. Figure~\ref{fig:infer_time} presents this comparison. Figure~\ref{fig:comp} plots average FAD against inference time across all baselines, where the ideal operating point lies in the lower-left corner (low FAD, fast inference). Video-Robin occupies this position, achieving the lowest average FAD and being faster than VidMuse, the only model with comparable audio quality. This demonstrates that Video-Robin's efficiency gains do not come at the cost of generation quality; A combination of patch-based autoregressive planning and an Euler-based diffusion solver enables both superior fidelity and substantially reduced inference time.

\subsection{Effect of Removing textual guidance}
We hypothesize that textual guidance is important in producing music that is semantically-aligned to the visual frame. To analyze this, we do a test-time comparison as follows. In the first case, we provide video frames and text prompt to the model for generation and the second case, we provide \textbf{only} the visual frames as input. Table \ref{tab:ablation-no-text} reports how adding text prompts provided a clear improvement across audio fidelity and diversity metrics.

\subsection{Human Evaluation}
\label{sec:human_eval}
We conducted A/B testing with 18 participants to measure Video-Robin's preference over baselines. These participants were aged 18 to 30 years, forming a realistic estimate of the general audience that consumes short form content like reels and YouTube shorts. We compared 20 videos across 6 models. Evaluators were presented with the same video paired with music outputs from 2 different models (selected at random) at a time and asked for their preference over 4 rubrics, resulting in 4 binary choices at every step. Each pair of video-music pairs was rated by 3 judges and we adopted the majority choice from each trio in our win-rate calculation. Below are the 4 major axes along which we collected user preference:

\begin{enumerate}
    \item \textbf{Audio Quality.} The perceived technical fidelity of the audio signal, assessed in terms of clarity, the absence of noise, distortion, clipping, and other sound artefacts, and the consistency of the recording throughout its duration. 
    \item \textbf{Musicality.} The intrinsic quality of the generated track as a standalone piece of music, judged independently of the accompanying video. This includes the coherence of melody and harmony, the naturalness of rhythm and structure, and the overall listenability and engagement of the music on its own terms.
    \item \textbf{Video-Music Alignment.} The degree to which the music corresponds to the video both temporally and semantically. Temporal correspondence refers to rhythmic synchronisation between musical beats or transitions and visual events such as cuts, motion, or scene changes. Semantic correspondence refers to whether the mood, genre, and emotional tone of the music are appropriate for the content depicted in the video.
    \item \textbf{Overall Assessment.} A holistic judgement of how well the music complements the video when all aspects are considered together, reflecting the participant's overall preference for one version over the other as a complete audiovisual experience.
\end{enumerate}

\begin{figure}
    \centering
    \includegraphics[width=0.8\linewidth]{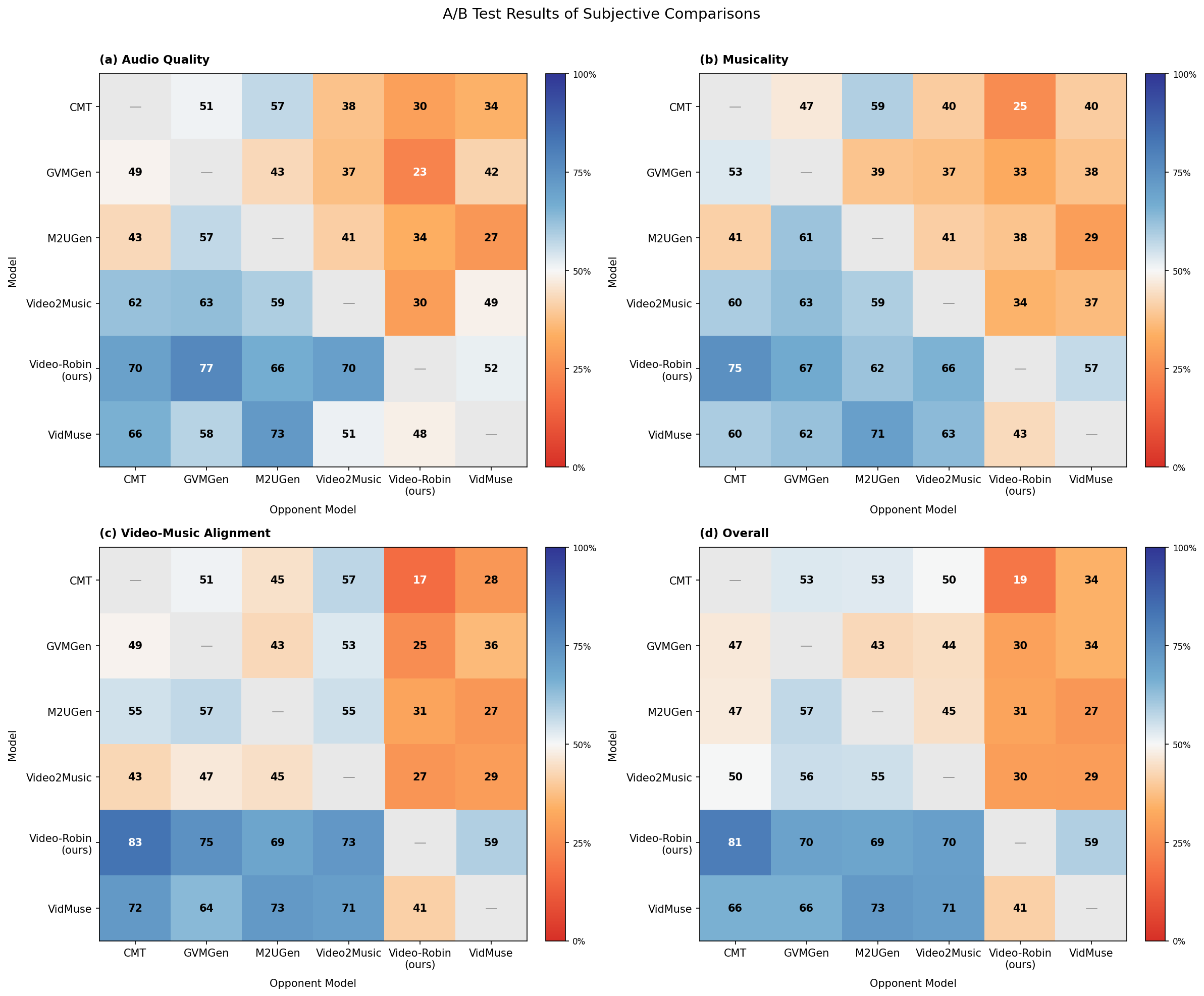}
    \caption{A/B Test results of the user study. The criteria used is discussed in the Section \ref{sec:human_eval}}
    \label{fig:human_eval}
\end{figure}


\begin{table}[H]
\centering
\caption{Results on the ReelBench, LORIS \cite{yu2023longtermrhythmicvideosoundtracker}, and V2MBench \cite{tian2025vidmusesimplevideotomusicgeneration} datasets. \textbf{Bold} indicates best and \underline{underline} indicates second-best among generated models.}
\label{tab:results}
\resizebox{\textwidth}{!}{%
\begin{tabular}{ll ccccccc}
\toprule
Dataset & Model & FAD ($\downarrow$) & FD ($\downarrow$) & KL ($\downarrow$) & IS ($\uparrow$) & IB ($\uparrow$) & Density ($\uparrow$) & Coverage ($\uparrow$) \\
\midrule
\multirow{7}{*}{ReelBench}
 & GT                                 & --      & --      & --     & --                   & 0.1417 & 0.9900 & 0.8800 \\
\cmidrule{2-9}
 & CMT \cite{Di_2021}                &  8.7522 & 37.7945 & 1.7329 & 1.2243 $\pm$ 0.0147 & \underline{0.1119} & 0.1084     & 0.0614 \\
 & GVMGen \cite{zuo2025gvmgengeneralvideotomusicgeneration}  &  3.5729 & 16.2638 & 1.5573 & \underline{1.7085 $\pm$ 0.0281} & 0.0957 & 0.0835 & 0.3881 \\
 & M2UGen \cite{liu2024mumullamamultimodalmusicunderstanding} &  4.5767 & 27.4208 & 1.5301 & 1.6499 $\pm$ 0.0567 & 0.0722 & 0.1094 & 0.2761 \\
 & Video2Music \cite{Kang_2024}      & 22.6459 & 73.0670 & 1.8839 & 1.0233 $\pm$ 0.0014 & 0.0473 & \textbf{0.1647} & 0.0084 \\
 & VidMuse \cite{tian2025vidmusesimplevideotomusicgeneration} & \underline{2.3022} & \underline{14.5385} & \underline{1.3194} & 1.4549 $\pm$ 0.0281 & \textbf{0.1233} & 0.1377 & \underline{0.5213} \\
 & Video-Robin (Ours)                 & \textbf{1.5110} & \textbf{10.9020} & \textbf{1.2556} & \textbf{2.0586 $\pm$ 0.0472} & 0.1017 & \underline{0.1384} & \textbf{0.5259} \\
\midrule
\multirow{7}{*}{LORIS \cite{yu2023longtermrhythmicvideosoundtracker}}
 & GT                                 & --      & --      & --     & --                   & 0.1558 & 0.5450 & 0.7550 \\
\cmidrule{2-9}
 & CMT \cite{Di_2021}                & 12.9733 & 37.3803 & 1.2515 & 1.2297 $\pm$ 0.0216 & \underline{0.0831} & 0.2133 & 0.0328 \\
 & GVMGen \cite{zuo2025gvmgengeneralvideotomusicgeneration}  & \underline{5.3595} & \textbf{17.8357} & \underline{1.2232} & \underline{1.7093 $\pm$ 0.0331} & 0.0771 & 0.2138 & \underline{0.1934} \\
 & M2UGen \cite{liu2024mumullamamultimodalmusicunderstanding} &  5.9096 & 28.0779 & \textbf{1.2203} & 1.6318 $\pm$ 0.0352 & 0.0694 & \textbf{0.4007} & 0.1852 \\
 & Video2Music \cite{Kang_2024}      & 31.6391 & 80.1407 & 1.2904 & 1.0071 $\pm$ 0.0007 & 0.0735 & 0.0596 & 0.0004 \\
 & VidMuse \cite{tian2025vidmusesimplevideotomusicgeneration} &  8.4983 & 34.4664 & 1.2800 & 1.2851 $\pm$ 0.0255 & \textbf{0.0878} & 0.2293 & 0.1259 \\
 & Video-Robin (Ours)                 & \textbf{4.1269} & \underline{27.6547} & 1.2431 & \textbf{2.0890 $\pm$ 0.1092} & 0.0821 & \underline{0.3094} & \textbf{0.2580} \\
\midrule
\multirow{7}{*}{V2MBench \cite{tian2025vidmusesimplevideotomusicgeneration}}
 & GT                                 & --      & --      & --     & --                   & 0.2474 & 0.6911 & 0.7775 \\
\cmidrule{2-9}
 & CMT \cite{Di_2021}                &  7.7565 & 41.6174 & 1.6732 & 1.2193 $\pm$ 0.0202 & 0.1590 & 0.4650 & 0.1942 \\
 & GVMGen \cite{zuo2025gvmgengeneralvideotomusicgeneration}  &  4.2146 & \underline{29.8336} & 1.6444 & \underline{1.5932 $\pm$ 0.1103} & 0.1952 & 0.3959 & 0.3585 \\
 & M2UGen \cite{liu2024mumullamamultimodalmusicunderstanding} &  5.5885 & 46.9329 & 1.8706 & 1.5799 $\pm$ 0.1545 & 0.1229 & \underline{0.4833} & 0.2372 \\
 & Video2Music \cite{Kang_2024}      & 29.8547 & 93.7820 & 2.0030 & 1.0054 $\pm$ 0.0009 & 0.0804 & 0.1255 & 0.0132 \\
 & VidMuse \cite{tian2025vidmusesimplevideotomusicgeneration} & \textbf{1.8577} & \textbf{22.4234} & \textbf{1.4039} & 1.4897 $\pm$ 0.0747 & \textbf{0.2280} & \textbf{0.6357} & \textbf{0.6205} \\
 & Video-Robin (Ours)                 & \underline{2.4264} & 32.3965 & \underline{1.6199} & \textbf{1.9097 $\pm$ 0.1139} & \underline{0.2082} & \underline{0.5835} & \underline{0.4512} \\
\bottomrule
\end{tabular}%
}
\end{table}

\begin{table}[H]
\centering
\caption{Audio-visual alignment evaluation results on ReelBench on video-to-music models that accept any additional features other than the video. \textbf{Bold} is best and \underline{underline} is second-best among generated models.}
\label{tab:av_results}
\resizebox{\textwidth}{!}{%
\begin{tabular}{l cccccc|c}
\toprule
Model & Rhythm ($\uparrow$) & Theme ($\uparrow$) & Emotion ($\uparrow$) & Culture ($\uparrow$) & Temporal ($\uparrow$) & Instrument Fit ($\uparrow$) & Overall ($\uparrow$) \\
\midrule
GT                                 & 2.906 & 2.977 & 3.084 & 3.602 & 2.612 & 3.258 & 2.860 \\
\cmidrule{1-8}
M2UGen \cite{liu2024mumullamamultimodalmusicunderstanding}      & 2.490          & 2.731          & \underline{3.000} & 3.255          & 2.143          & \underline{3.100} & \underline{2.724} \\
Video2Music \cite{Kang_2024}      & \underline{2.622} & \underline{2.744} & 2.744          & \underline{3.293} & \underline{2.268} & 3.090          & 2.677          \\
Video-Robin (Ours)                 & \textbf{2.786} & \textbf{2.890} & \textbf{3.100} & \textbf{3.565} & \textbf{2.472} & \textbf{3.200} & \textbf{2.753} \\
\bottomrule
\end{tabular}%
}
\end{table}

\begin{table}[H]
\centering
\caption{\textbf{Detailed ablation to assess the effect of removing FSQ \& RITE.} We provide results across multiple datasets (ReelBench, LORIS \cite{yu2023longtermrhythmicvideosoundtracker} and V2MBench \cite{tian2025vidmusesimplevideotomusicgeneration}) to show how these trends generalise across different datasets. \textbf{Bold} is best and \underline{underline} is second-best among generated models.}
\label{tab:ablation-fsq-rite}
\resizebox{\textwidth}{!}{%
\begin{tabular}{ll ccccccc}
\toprule
Dataset & Model & FAD ($\downarrow$) & FD ($\downarrow$) & KL ($\downarrow$) & IS ($\uparrow$) & IB ($\uparrow$) & Density ($\uparrow$) & Coverage ($\uparrow$) \\
\midrule
\multirow{2}{*}{ReelBench}
 & Video-Robin   & \textbf{1.5110} & \textbf{10.9020} & \textbf{1.2556} & \textbf{2.0586 $\pm$ 0.0472} & \textbf{0.1017} & \textbf{0.1384} & \textbf{0.5259} \\
 
 & w/o RITE   & 6.599 & 46.5526 & 1.1953 & 1.1337 $\pm$ 0.0078 & 0.1098 & 0.0367 & 0.137 \\
 & w/o FSQ + RITE        &         4.3501  &         25.8103  &         1.4291  &         1.1946 $\pm$ 0.0208  &         0.0617  &         0.0678  &         0.1741  \\
\midrule
\multirow{2}{*}{LORIS \cite{yu2023longtermrhythmicvideosoundtracker}}
 & Video-Robin     & \textbf{4.1269} & \textbf{27.6547} & 1.2431 & \textbf{2.0890 $\pm$ 0.1092} & \textbf{0.0821} & \textbf{0.3094} & \textbf{0.2580} \\
 & w/o RITE        &         11.95  &         61.0521  &   \textbf{1.2091}  &     1.1254 $\pm$ 0.0059        &         0.0175  &         0.0725 & 0.0213  \\
 & w/o FSQ + RITE        &         7.9836  &         38.0861  &         1.2530  &         1.1889 $\pm$ 0.0136  &         0.0734  &         0.1947  &         0.1099  \\
\midrule
\multirow{2}{*}{V2MBench \cite{tian2025vidmusesimplevideotomusicgeneration}}
 & Video-Robin    & \textbf{2.4264} & \textbf{32.3965} & 1.6199 & \textbf{1.9097 $\pm$ 0.1139} & \textbf{0.2082} & \textbf{0.5835} & \textbf{0.4512} \\
 & w/o RITE        &  8.6789  &  65.8347  &       \textbf{1.4963}  &    1.1214 $\pm$ 0.0233  &         0.1455  &    0.1402 &0.0705 \\
 & w/o FSQ + RITE        &         6.6984  &         48.9303  &         1.6206  &         1.1877 $\pm$ 0.0263  &         0.1916  &         0.3919  &         0.1996  \\
\bottomrule
\end{tabular}%
}
\end{table}

\begin{table}[H]
\centering
\caption{\textbf{Ablation to assess the effect of patch size} on model performance across ReelBench, LORIS \cite{yu2023longtermrhythmicvideosoundtracker} and V2MBench \cite{tian2025vidmusesimplevideotomusicgeneration}. \textbf{Bold} is best and \underline{underline} is second-best among generated models.}
\label{tab:ablation-patch}
\resizebox{\textwidth}{!}{%
\begin{tabular}{ll ccccccc}
\toprule
Dataset & Patch Size & FAD ($\downarrow$) & FD ($\downarrow$) & KL ($\downarrow$) & IS ($\uparrow$) & IB ($\uparrow$) & Density ($\uparrow$) & Coverage ($\uparrow$) \\
\midrule
\multirow{2}{*}{ReelBench}
 & 4   & \textbf{1.5110} & \textbf{10.9020} & \textbf{1.2556} & \textbf{2.0586 $\pm$ 0.0472} & \textbf{0.1017} & \textbf{0.1384} & \textbf{0.5259} \\
 & 8 & 4.1271   &         12.6692  &         1.4253  &         1.5859 $\pm$ 0.0265  &  0.0579  &    0.0894       &    0.1724  \\
 & 16        & 3.779  &   16.2368  &         1.4069  &         1.4091 $\pm$ 0.0263  &  0.0660  &         0.1305  &   0.1405  \\
\midrule
\multirow{2}{*}{LORIS \cite{yu2023longtermrhythmicvideosoundtracker}}
 & 4     & \textbf{4.1269} & \textbf{27.6547} & 1.2431 & \textbf{2.0890 $\pm$ 0.1092} & \textbf{0.0821} & 0.3094 & 0.2580 \\
 & 8        &  3.062  &  18.1955  &  \textbf{1.203}  &   1.5438 $\pm$ 0.0392  &  0.0683  &   \textbf{0.4531} & \textbf{0.3969} \\
 & 16        &         5.6008  &  25.8981  &  1.3478  &   1.4172 $\pm$ 0.0320   &   0.0597  &   0.2845 & 0.2671  \\
\midrule
\multirow{2}{*}{V2MBench \cite{tian2025vidmusesimplevideotomusicgeneration}}
 & 4   & \textbf{2.4264} & \textbf{32.3965} & 1.6199 & \textbf{1.9097 $\pm$ 0.1139} & \textbf{0.2082} & \textbf{0.5835} & \textbf{0.4512} \\
 & 8 & 4.7606  & 31.511 &   \textbf{1.5343}  &   1.4238$\pm$ 0.0865  & 0.1277  &   0.5278 & 0.3973 \\
 & 16        & 7.2063 & 33.9057  &  1.6908 & 1.3766 $\pm$ 0.0596 &  0.1384  & 0.3364 & 0.3248 \\
\bottomrule
\end{tabular}%
}
\end{table}

\begin{table}[H]
\centering
\caption{\textbf{Ablation to assess the effect of removing textual guidance} across ReelBench, LORIS \cite{yu2023longtermrhythmicvideosoundtracker} and V2MBench \cite{tian2025vidmusesimplevideotomusicgeneration}. \textbf{Bold} is best and \underline{underline} is second-best among generated models.}
\label{tab:ablation-no-text}
\resizebox{\textwidth}{!}{%
\begin{tabular}{ll ccccccc}
\toprule
Dataset & Model & FAD ($\downarrow$) & FD ($\downarrow$) & KL ($\downarrow$) & IS ($\uparrow$) & IB ($\uparrow$) & Density ($\uparrow$) & Coverage ($\uparrow$) \\
\midrule
\multirow{2}{*}{ReelBench}
 & Video-Robin   & \textbf{1.5110} & \textbf{10.9020} & \textbf{1.2556} & \textbf{2.0586 $\pm$ 0.0472} & \textbf{0.1017} & \textbf{0.1384} & \textbf{0.5259} \\
& w/o text prompt & 2.9019 & 25.4968 & 1.4678 & 1.9054 $\pm$ 0.0413 & 0.0799 & 0.0510 & 0.3470 \\
\midrule
\multirow{2}{*}{LORIS \cite{yu2023longtermrhythmicvideosoundtracker}}
 & Video-Robin     & \textbf{4.1269} & \textbf{27.6547} & \textbf{1.2431} & \textbf{2.0890 $\pm$ 0.1092} & \textbf{0.0821} & \textbf{0.3094} & \textbf{0.2580} \\
& w/o text prompt & 4.5609 & 30.9807 & 1.2770 & 1.8495 $\pm$ 0.0843 & 0.0597 & 0.2972 & 0.2511 \\

\midrule
\multirow{2}{*}{V2MBench \cite{tian2025vidmusesimplevideotomusicgeneration}}
 & Video-Robin    & \textbf{2.4264} & \textbf{32.3965} & 1.6199 & \textbf{1.9097 $\pm$ 0.1139} & \textbf{0.2082} & \textbf{0.5835} & \textbf{0.4512} \\
 & w/o text prompt & 3.1152 & 38.3150 & \textbf{1.5408} & 1.5737 $\pm$ 0.2212 & 0.1774 & 0.5728 & 0.2946 \\
\bottomrule
\end{tabular}
}
\end{table}

\section{Conclusion, Limitations and Future Work}

We introduce Video-Robin, a text and video-conditioned video-to-music generation framework that explicitly models creator intent while preserving audiovisual alignment. The key idea is to factor generation into (i) autoregressive multimodal planning over continuous latent patches (via a SemanticLM + FSQ bottleneck + residual integration) and (ii) diffusion-based refinement that converts each planned patch into high-fidelity music latents, which are finally decoded by a pretrained VAE. This hybrid design addresses the long-standing trade-off between global musical structure and local acoustic fidelity, enabling controllable synthesis without sacrificing realism. Empirically, Video-Robin achieves strong audio quality, diversity, and alignment across both in-domain and out-of-domain benchmarks, while also delivering substantially faster inference.\\
Video-Robin currently targets a constrained setting--our experiments use 10-second clips and instrumental/background music only--which does not yet capture long-form scoring needs like multi-minute narrative arcs, motif recurrence, or seamless transitions across scene cuts; extending the planner–refiner hierarchy to longer videos with cut/beat-aware planning and timestamped “hit-point” controls is an immediate next step. Evaluation is also imperfect: while we report standard audio quality/diversity metrics and an ImageBind-based alignment proxy, music–video fit and creator intent-following remain hard to measure reliably, motivating music-centric alignment metrics (e.g., rhythmic hit accuracy, pacing/structure match) and human-preference supervision for controllability. Finally, the system depends on frozen representation components (e.g., VAE latent space / pretrained encoders) and a relatively small intent-annotated set (ReelBench), which may cap expressivity in niche genres and limit coverage of real creator workflows; scaling intent-rich data, adding interactive editing (inpainting/partial re-scoring), and exploring more adaptive or learned latents/encoders are concrete directions to broaden applicability and improve fidelity–control trade-offs.



%
%
\bibliographystyle{splncs04}
\bibliography{main}

\newpage
\section*{Appendix A: Alternative for Audiovisual Alignment Evaluation}
\label{sec:gemini_eval}

Recent advances in multimodal foundation models enable the use of large language models as \textit{Omni-Judges} for evaluating generative systems. While models such as ImageBind learn shared embeddings across modalities, they are not explicitly trained to assess fine-grained music-video alignment. As a result, embedding similarity alone may not capture temporal, emotional, or structural coherence between generated music and visual content. To address this limitation, we employ Gemini as an Omni-Judge to perform structured audio-visual coherence evaluation.\\
Gemini is provided with a single video file containing both the visual content and the generated music track. When available, we also supply the text prompt used for music generation. The model is instructed through carefully designed system and evaluation prompts to act as an expert audio-visual evaluator and produce a strictly formatted JSON response. The evaluation begins with a short global analysis describing the relationship between the visual pacing and mood of the video and the musical structure and affect of the generated audio. The model then assigns quantitative ratings across several predefined axes using a discrete 1-5 scale, where 1 indicates a severe mismatch and 5 indicates highly synchronised and coherent alignment.\\
The evaluation axes are defined as follows:

\begin{enumerate}

\item \textbf{Rhythmic Sync.} 
This axis measures whether the beat structure, tempo, and rhythmic patterns of the music align with the motion dynamics, visual cuts, and overall pacing of the video. It captures the degree of low-level temporal synchronization between musical rhythm and visual events.

\item \textbf{Theme Coherence.} 
This axis evaluates whether the genre and stylistic characteristics of the music are appropriate for the subject matter and setting depicted in the video. Gemini also predicts dominant single-word theme labels for both the video and the audio (e.g., \textit{nature}, \textit{urban}, \textit{scifi}) and determines whether the thematic alignment between the modalities is meaningful.

\item \textbf{Emotion Alignment.} 
This axis assesses whether the emotional tone conveyed by the music corresponds to the visual mood of the video. The model predicts dominant emotional labels for both modalities (e.g., \textit{happy}, \textit{tense}, \textit{melancholic}, \textit{calm}) and indicates whether the emotional correspondence between the audio and visuals is consistent.

\item \textbf{Cultural Relevance.} 
This axis measures whether the instrumentation, musical style, and sonic motifs are culturally appropriate for the context shown in the video. For example, videos depicting culturally specific settings may require regionally relevant musical styles or instrumentation.

\item \textbf{Temporal Dynamics.} 
This axis evaluates whether structural changes in the music, such as builds, drops, transitions, or fades, align with important visual events including scene changes, camera movements, or climactic moments in the video.

\item \textbf{Instrumentation Fit.} 
This axis measures whether the specific instruments, synthesizers, and timbral characteristics used in the generated music are suitable for the visual aesthetic and narrative context of the video.

\item \textbf{Overall Alignment.} 
This axis captures the holistic integration of the audio and visual streams, reflecting how naturally the generated music complements the video when considering all aspects of temporal, thematic, emotional, and structural correspondence.

\end{enumerate}

\begin{figure}
    \centering
    \includegraphics[width=\linewidth]{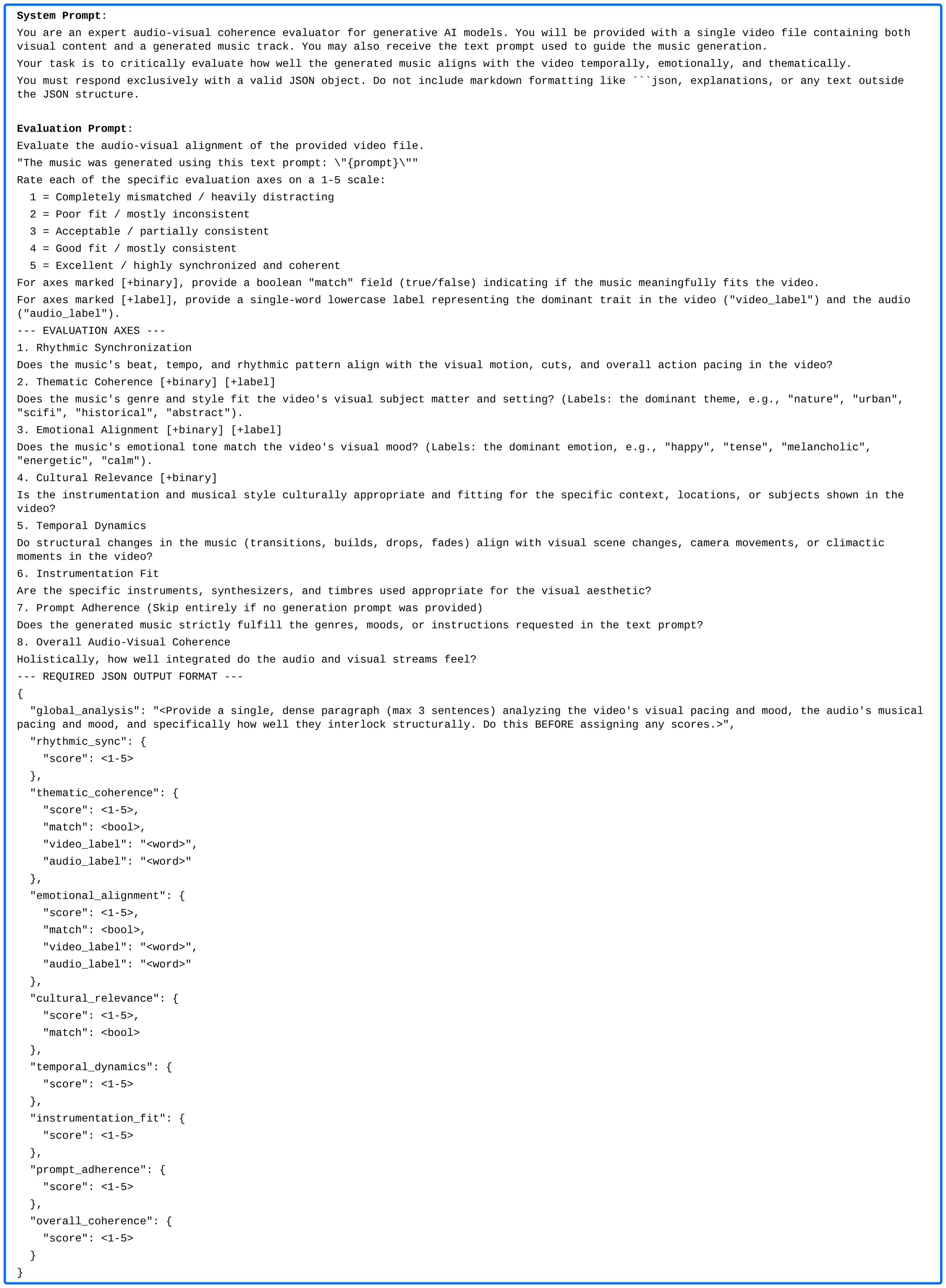}
    \caption{System prompt and evaluation prompt used to configure Gemini as an Omni-Judge for audio-visual alignment evaluation.}
    \label{fig:gemini_prompt}
\end{figure}

Figure~\ref{fig:gemini_prompt} presents the complete prompt used to configure Gemini as an Omni-Judge for audio-visual coherence evaluation.

\section*{Appendix B: Dataset Preprocessing prompts}
\begin{figure}
    \centering
    \includegraphics[width=\linewidth]{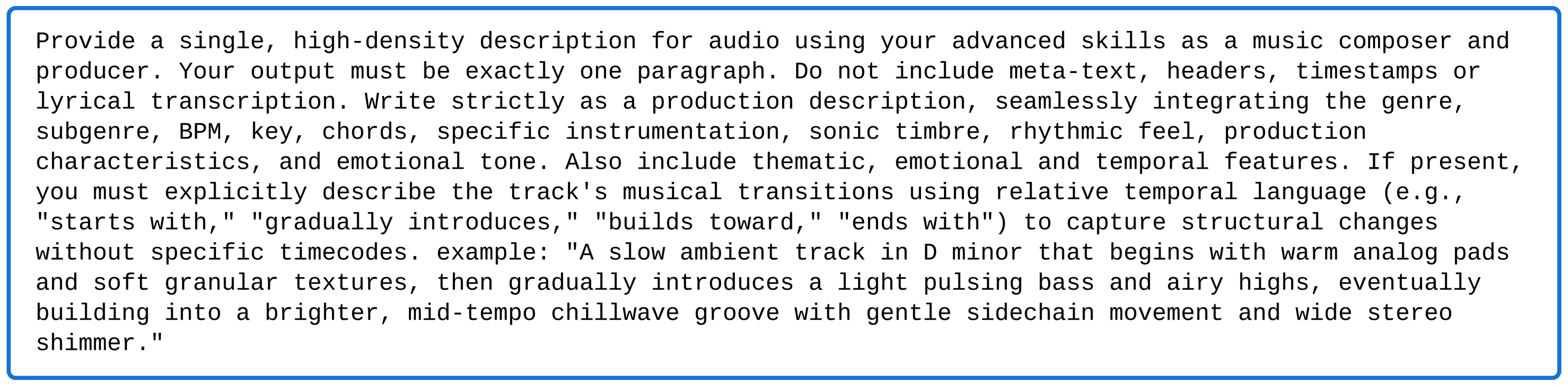}
    \caption{Prompt provided to MusicFlamingo for extracting fine-grained musical attributes from the ground-truth audio track.}
    \label{fig:mf_prompt}
\end{figure}

\begin{figure}
    \centering
    \includegraphics[width=\linewidth]{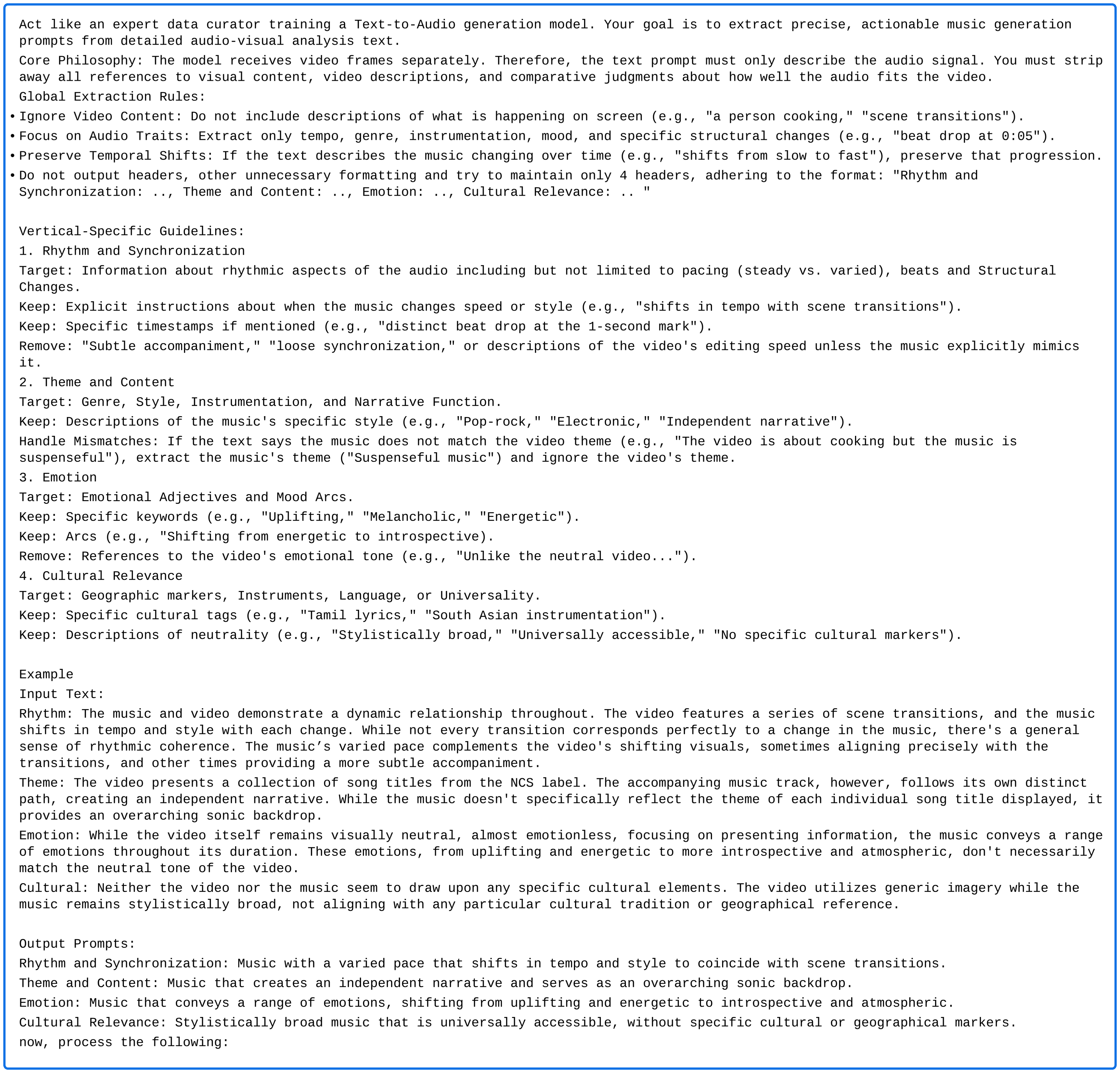}
    \caption{Prompt provided to Qwen3-8B for paraphrasing Harmony dataset captions into music generation prompts.}
    \label{fig:qwen_harmony}
\end{figure}

\begin{figure}
    \centering
    \includegraphics[width=\linewidth]{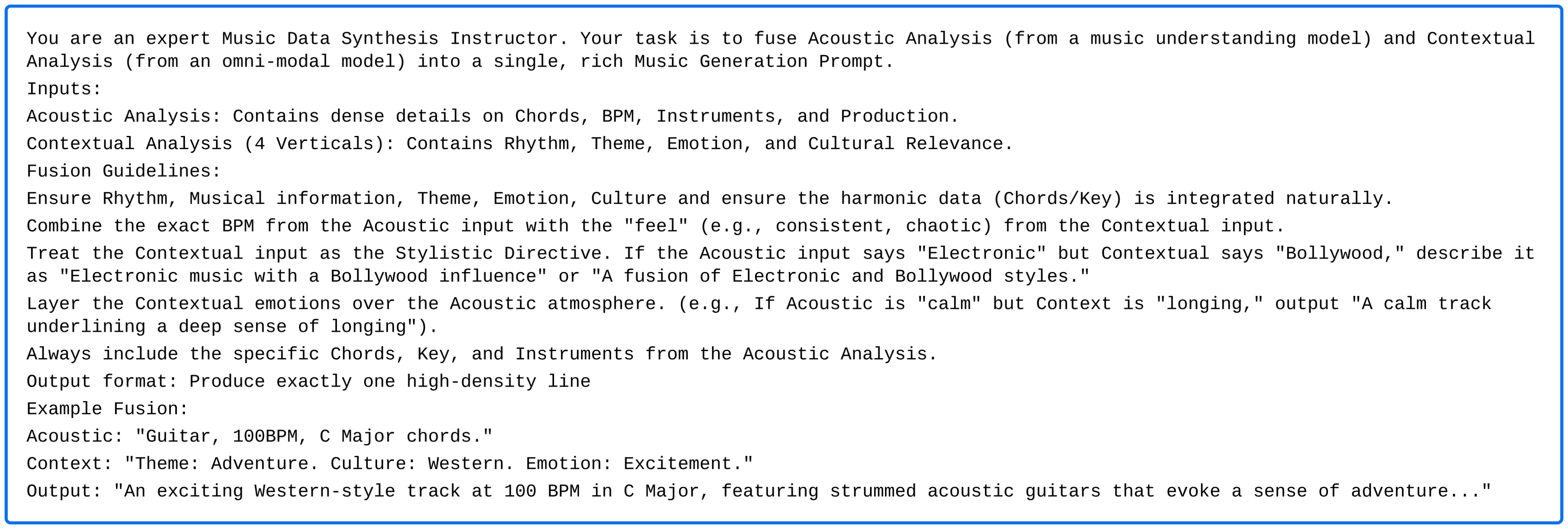}
    \caption{Prompt provided to Qwen3-8B for combining MusicFlamingo-extracted musical attributes with the paraphrased Harmony prompt to produce the final training prompt.}
    \label{fig:qwen_comb}
\end{figure}

As detailed in Section~\ref{sec:dataset}, we employ a multi-step data preprocessing pipeline to construct the training dataset. Figure~\ref{fig:mf_prompt} shows the prompt supplied to MusicFlamingo to extract fine-grained information from the ground-truth music file. We then provide the prompt shown in Figure~\ref{fig:qwen_harmony} to Qwen3-8B to paraphrase the caption from the Harmony dataset into a music generation prompt. Finally, we prompt Qwen3-8B to combine the fine-grained information obtained from MusicFlamingo with the paraphrased Harmony dataset prompt using the prompt shown in Figure~\ref{fig:qwen_comb}.

\begin{figure}[H]
    \centering
    \includegraphics[width=0.9\linewidth]{wtch_analysis.png}
    \caption{\textbf{Qualitative spectrogram comparison on a meditative ambient-drone sample.} Video-Robin best reproduces active high-frequency content (6–8 kHz) consistent with the "crystalline bells and airy whistles" in the prompt.}
    \label{fig:wtch}
\end{figure}

\begin{figure}[H]
    \centering
    \includegraphics[width=0.9\linewidth]{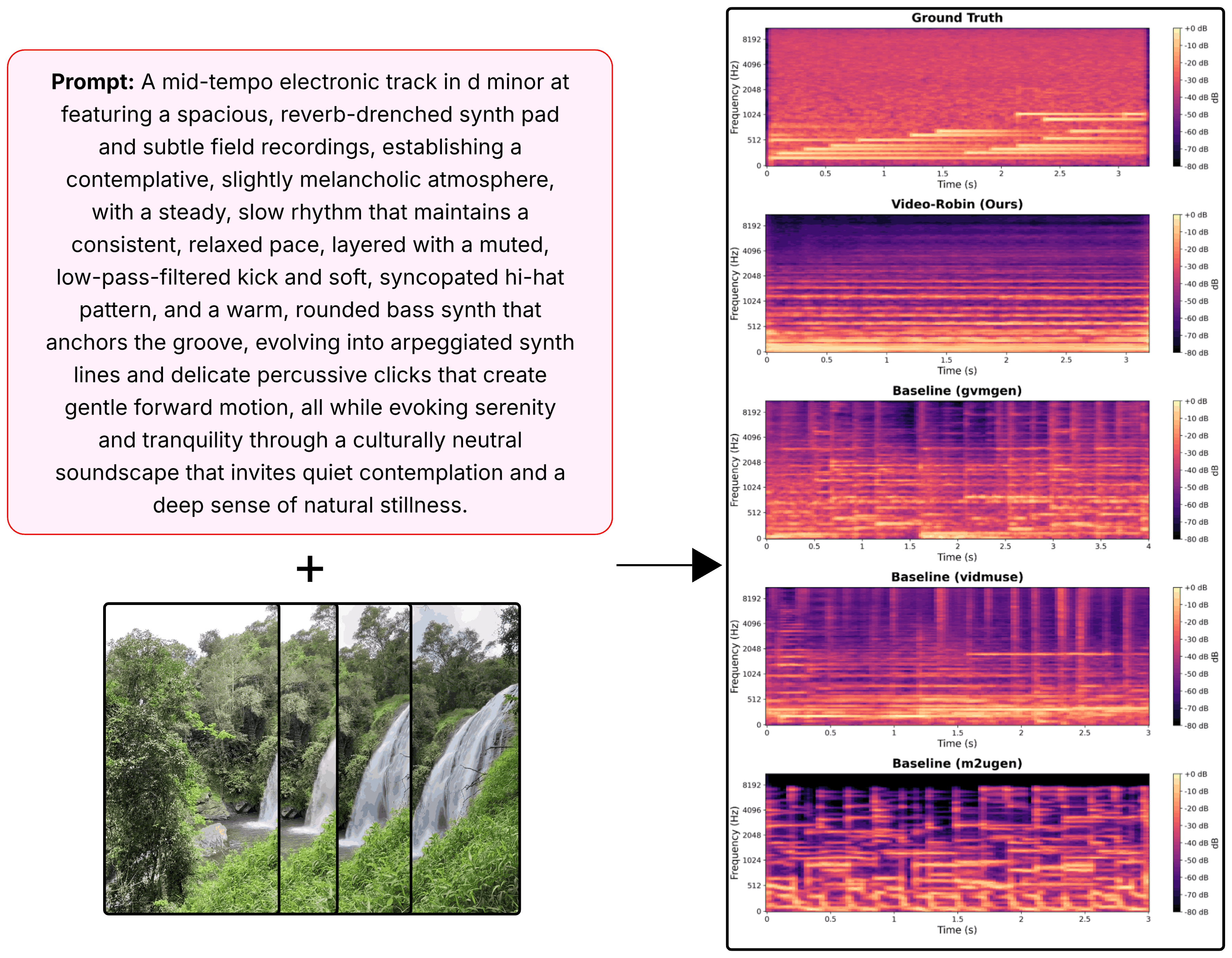}
    \caption{\textbf{Qualitative spectrogram comparison on a contemplative mid-tempo electronic sample.} Video-Robin produces clear horizontal banding in the mid-to-low frequencies consistent with the "reverb-drenched synth pad and warm rounded bass synth" in the prompt, while appropriately attenuating high-frequency energy in line with the prompt's "muted, low-pass-filtered" character.}
    \label{fig:b29}
\end{figure}
\section*{Appendix C: Qualitative Analysis of Spectrograms}
Figures \ref{fig:wtch} and \ref{fig:b29} compare spectrograms of outputs generated by different baselines for two input examples, alongside results from our model, Video-Robin. We analyze three key aspects of the generated audio below.\\
\textbf{Spectral Coverage.}
Video-Robin produces the broadest frequency coverage among the evaluated models when conditioned on text prompts, closely matching the ground-truth energy distribution. In contrast, GVMGen tends to attenuate high-frequency components, while M2UGen exhibits a noticeable cutoff around 8 kHz. VidMuse occasionally generates excessive high-frequency energy, leading to localized saturation. By comparison, Video-Robin more consistently captures both low–mid frequency structure and higher-frequency detail associated with the target music.\\
\textbf{Temporal Consistency.}
Video-Robin generally maintains energy throughout the duration of the generated clip, indicating stable temporal generation. However, we observe occasional mid-clip energy drops in some samples, which can lead to partial temporal collapse. In contrast, GVMGen, VidMuse, and M2UGen typically produce temporally continuous outputs, although their spectral coverage is more limited.\\
\textbf{Prompt Adherence.}
Text conditioning enables Video-Robin to better reflect fine-grained musical attributes specified in prompts. For example, prompts describing low-pass filtered textures correspond to reduced high-frequency energy, ambient prompts yield sustained pad-like structures, and energetic prompts produce denser transient patterns. In contrast, the baseline models, which rely primarily on visual inputs, tend to generate stylistically plausible but less prompt-specific audio outputs.

\end{document}